\documentclass[aps,twocolumn,showpacs,preprintnumbers,amssymb]{revtex4}

\newcommand{\be}{\begin{eqnarray}}
\newcommand{\ee}{\end{eqnarray}}

\newcommand{\eins}{\mbox{$1 \hspace{-1.0mm}  {\bf l}$}}
\def\bea{\begin{eqnarray}}
\def\eea{\end{eqnarray}}
\def\C{\hbox{$\mit I$\kern-.7em$\mit C$}}
\def\N{\hbox{$\mit I$\kern-.3em$\mit N$}}

\def\l{\langle}
\def\r{\rangle}

\usepackage{dcolumn} 
\usepackage{bm} 

\usepackage{epsfig}

\begin{document}

\title{Entanglement purification for Quantum Computation}

\author{W. D\"{u}r  and H.-J. Briegel}

\affiliation{
Sektion Physik, Ludwig-Maximilians-Universit\"at M\"unchen, Theresienstr.\ 37, D-80333 M\"unchen, Germany.}

\date{\today}

\begin{abstract}
We show that thresholds for fault--tolerant quantum computation are solely determined by the quality of single--system operations if one allows for $d$--dimensional systems with $8 \leq d \leq 32$. Each system serves to store one logical qubit 
and additional auxiliary dimensions are used to create and purify entanglement between systems. Physical, possibly probabilistic two--system operations with error rates up to $2/3$ are still tolerable to realize deterministic high quality two--qubit gates on the logical qubits. The achievable error rate is of the same order of magnitude as of the single--system operations. We investigate possible implementations of our scheme for several physical set--ups.
\end{abstract}


\maketitle


Much of the theoretical and experimental interest in quantum information theory in the last decade has been devoted to quantum computation. The finding of quantum algorithms which offer an (exponential) speedup over their best known classical counterparts \cite{Sh96} as well as the possibility to operate a quantum computer in a noisy environment in a fault--tolerant way \cite{faulttolerant} can be counted as milestones of this investigation.  Since then, many theoretical proposals to implement quantum computation in various physical systems, ranging from trapped atoms or ions to NMR and quantum dots, have been put forward and experimental implementation of basic quantum logic gates was demonstrated in several of these systems \cite{Fo00}. Unfortunately, there are stringent requirements which have to be fulfilled before a universal quantum computer can operate in a fault tolerant way. These include gate error rates below a threshold value which is of the order of $10^{-4} - 10^{-5}$, still far beyond experimentally reachable accuracies. On the other hand, in quantum communication it was found \cite{Br98, Gi99} that the requirements to ensure secure communication over arbitrary distances are much less stringent. Indeed, error rates of the order of several percent are tolerable in this case \cite{Br98,Gi99}. The main tool to achieve secure  \cite{As99} quantum communication over arbitrary distances is {\it entanglement purification} \cite{Be96}.
But is entanglement purification also useful for quantum computation? Does it allow one to increase thresholds for tolerable errors? In this letter, we will answer these questions in a positive way. We will show that one can indeed use entanglement purification 
to increase the quality of two--system operations by several orders of magnitude. This in turn implies that {\it the requirements for fault  tolerant quantum computation can be met if the quality of single--system operations is sufficiently high}, (almost) independently of the quality of two--system operations.


We consider a collection of distinct physical systems which shall be used to perform a quantum computation. Each of the systems serves to store at least one qubit of information. That such a set--up can be used for fault tolerant quantum computation requires --among others-- the ability to perform both arbitrary unitary operation on each of the distinct systems and controlled interactions between different systems (i.e. non--local operations) with error rates below $10^{-4}-10^{-5}$ \cite{Ah02}. 
However, not all of these operations are equally difficult to perform. For example, it may be easy to manipulate each of the distinct systems in a controlled way while interactions between different systems may be very difficult to achieve. Consider the example where each system corresponds to the polarization degrees of freedom of a photon. While the state of each photon may be manipulated quite easily by means of linear optical elements, controlled deterministic interactions between photons (e.g. using Kerr non--linearities) are very difficult to achieve. Also for trapped neutral atoms or ions, it is much easier to manipulate the electronic states of each particle by means of well controllable laser pulses than to achieve a controlled interaction between two particles. 

Having already initiated the discussion with atoms and ions, we will refer to each physical system as ``particle''. It is however not a necessary requirement of our analysis that each distinct systems corresponds to a real particle, it could also be some abstract system. 
In what follows, we will carefully distinguish between operations on a single particle and operations which require controlled interaction between two particles. Our results are applicable to all situations where single--particle operations are much easier to implement than two--particle operations.

In order to simplify the description and discussion of our scheme, we impose a virtual tensor product structure, that is we divide each physical $d$--level system (particle) into $n$ virtual qubit--subsystems \cite{Za01}, i.e. $d=2^n$. We will refer to different particles as $A, B, C, \ldots ,Z$, while virtual subsystems of, say, particle $A$ are denoted by $A_1,A_2,A_3,\ldots, A_n$.  The corresponding Hilbert space is denoted by ${\cal H}={\cal H}_A\otimes{\cal H}_B \otimes \ldots {\cal H}_Z$ with ${\cal H}_X={\cal H}_{X_1}\otimes{\cal H}_{X_2}\otimes \ldots {\cal H}_{X_n} \cong (\C^2)^{\otimes n}$, $ X\in\{A,B,\ldots, Z\}$ . 


A brief summary of the scheme is as follows.
Each particle $X$ serves to store and manipulate one logical qubit in its virtual subsystem $X_1$.
A (noisy) two--particle interaction is used to create entanglement between the additional virtual subsystems $X_k, k\geq 2$ of different particles. This noisy entanglement is efficiently purified using a novel entanglement purification scheme based on nested entanglement pumping which requires less than five virtual subsystems, i.e. $d \leq 32$. The entanglement is then used ---together with high--quality single--particle operations--- to implement in a deterministic way two--particle gates between the logical qubits. For instance, a CNOT-gate \cite{noteCNOT} between $A_1$ and $B_1$ can be realized using schemes presented in Ref. \cite{Go98,Go99,Ci00}. We find that the physical two--particle gate need only be weakly entangling or may even be probabilistic (i.e. the operation only needs to be successful with some non--zero probability) and the error rate can be as high as $2/3$. This still allows one to realize deterministic logical two--qubit gates whose quality is of the same order of magnitude as the single--particle operations. This means that the thresholds for fault tolerant quantum computation are solely determined by the quality of {\it single}--particle operations. The requirements to build a scalable quantum computer thus reduce to provide small ($d \leq 32$), well controllable systems which interact by some means, where the interaction may be very noisy or even probabilistic. In what follows, we will discuss this scheme in detail for two particles, $A$ and $B$. 


We start with the creation and purification of noisy entanglement. 
Consider a situation where several, say $n_0$, (noisy) entangled states shared between systems $A_k$ and $B_k$, $k\geq 2$, have been created using the physical two--particle interaction. 
Standard entanglement purification methods, e.g. the recurrence protocol of Ref. \cite{Be96}, can be applied to purify the $n_0$ noisy entangled pairs and eventually to end up with a single entangled pair of higher quality shared between $A_2$ and $B_2$. 
Imperfections in single--particle operations still allow us to increase the quality of the entangled pairs up to a certain point ---depending of the quality of single--particle operations--- however no maximally entangled states can be created under these circumstances \cite{Br98,Gi99}. Nevertheless, we will consider such situation in the following. Using the standard recurrence methods \cite{Be96,De96} (or other methods such as hashing or breeding \cite{Be96}) typically requires the storage of hundreds of entangled pairs. However, storing $n_0$ entangled pairs plus the logical qubits requires $d=2^{n_0+1}$ levels for each particle, which quickly becomes impractical simply because no sufficient number of controllable levels are available. 
To avoid this exponential overhead in the number of dimensions, we propose to use a novel, modified entanglement purification scheme which consists of {\it nested entanglement pumping}. 


The main idea is to use this entanglement pumping, that is to use the (noisy) two--particle gate to repeatedly create noisy entangled pairs between systems $A_2,B_2$, which are used to purify another pair shared between systems $A_3,B_3$ \cite{Br98}. The advantage of such scheme is that only two virtual subsystems per particle are required. However, as the purification process has to be restarted from the beginning as soon as one purification step fails, the time required to implement entanglement pumping as compared to the standard recurrence method is (polynomially) higher. It should be mentioned that even under ideal conditions, no maximally entangled states can be created using entanglement pumping, but the fidelity of the pairs can only be increased by a certain amount. This last problem can be overcome by using a nested scheme in such a way that the pair stored in $A_3,B_3$ is purified (almost) up to its highest reachable value and then used to purify another such pair (which was created in the same way) stored in systems $A_4,B_4$. For each nesting level, one additional virtual subsystem per particle is required. We have performed numerical investigation of the nested entanglement pumping scheme and found that when considering imperfect operations, 
the minimal required fidelity as well as the reachable fidelity of the pairs is the same as in the recurrence method of Ref. \cite{De96}, and only a few nesting levels are required \cite{Du02}. 
For all practical purposes, that is when the error rates for for single particle operations are above $10^{-7}$, three nesting levels are sufficient (i.e. a total of $d\leq32$ dimensions per particle). In this case, the achievable error rate of the {\it logical} two--qubit gate is of the same order of magnitude as the error rate of single--particle operations, provided that error rates for {\it physical} two--particle gates are at the order of $0.2$ or lower.
That is, the nested entanglement pumping combines the high tolerable error rates with few physical resources at the price of (polynomial) time overhead.   
Note that since entanglement pumping is itself a probabilistic process, the creation of the entangled pairs (and thus the two--particle gate) do not need to be deterministic. It only has to be known when the gate was successful. While such a probabilistic gate may completely destroy the performance of a quantum computation when applied directly to data--qubits (as the whole computation has to be restarted each time a gate fails), this is not the case in our proposal, since the information carrying qubits are uneffected  by the probabilistic gate. 


The purified entangled pair created in this way can then be used to implement {\it deterministically} a two--particle gate (e.g. a CNOT) between the logical qubits stored in $A_1$ and $B_1$. In case the pair is maximally entangled and the single--particle operations are perfect, this can be accomplished with unit fidelity. Given a maximally entangled state shared between $A_2,B_2$, $|\Phi\r\equiv 1/\sqrt{2}(|00\r+|11\r)$, the following sequence of single--particle operations realizes a CNOT--gate between $A_1,B_1$ \cite{Go98}: (i) CNOT$_{A_1,A_2}$, (ii) measurement of $A_2$ in $z$--basis, depending on outcome of measurement, apply $\eins_{B_2}$ (outcome '0') or $\sigma_x^{B_2}$ (outcome '1'), (iii) CNOT$_{B_2,B_1}$, (iv) measurement of $B_2$ in $x$--basis, depending on outcome of measurement, apply $\eins_{A_1}$ (outcome '0') or $\sigma_z^{A_1}$ (outcome '1').
In a similar way one may also realize arbitrary two--qubit operations (instead of CNOT) or even multi--qubit operation by purifying and consuming certain (multipartite) entangled states, following e.g. the scheme proposed in Ref. \cite{Ci00}. 
For non--maximally entangled pairs and imperfect single--particle operations, the two--particle gate is realized only in an imperfect way. The corresponding completely positive map ${\cal E}$ can be obtained by carrying out (i)-(iii), taking imperfections of single--particle gates and measurement into account, and considering a non--maximally entangled mixed state, e.g. of Werner form, $\rho_{A_2B_2}=x|\Phi\r\l\Phi|+(1-x)/4\eins$, which can always be achieved using depolarization. 
The average gate fidelity,
\be
{\bar F}({\cal E},U_{\rm CNOT})\equiv\int d\psi \l\psi|U_{\rm CNOT}^\dagger {\cal E}(\psi) U_{\rm CNOT} |\psi\r,\label{Faver}
\ee
is used as measure of the quality of the imperfect operation ${\cal E}$ and we refer to $p \equiv (1-{\bar F})$ as {\it error rate of the operation}. Note that given ${\cal E}$, ${\bar F}$ can be easily evaluated using the results of Ref. \cite{Ni02}.

\begin{figure}[ht]
\begin{picture}(230,100)
\put(-5,-5){\epsfxsize=230pt\epsffile[20 154 1057 620]{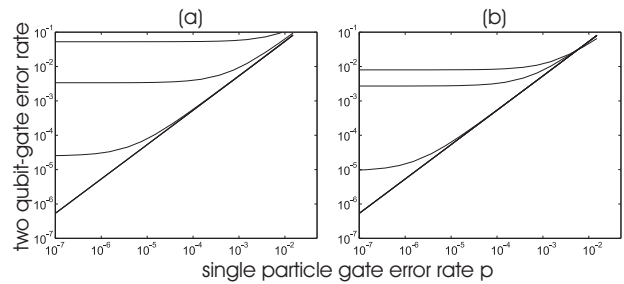}}
\end{picture}
\caption[]{Double logarithmic plot of achievable logical two--qubit gate error rate against single--particle error rate $p$ for fixed error rate of physical two--particle interaction of (a) $1.5*10^{-1}$, (b) $10^{-2}$. Curves from top to bottom correspond to no entanglement purification, entanglement pumping using 1, 2, 3 (or more) nesting levels respectively.}
\label{Fig1}
\end{figure}

We have analyzed the influence of imperfections on the scheme described above. In order to illustrate our results, we describe imperfect single--particle operations by a simple error model 
, however the application of our scheme is not restricted to such error model but universal. 
We describe imperfect single--particle operations acting on two virtual subsystems as follows
\be
{\cal E}_{U_{A_1A_2}}(\rho)=qU_{A_1A_2}\rho U_{A_1A_2}^\dagger+\frac{1-q}{4}\eins_{A_1A_2} \otimes {\rm tr}_{A_1A_2}\rho\label{noisyU}.
\ee
While with probability $q$ the desired gate is performed, with probability $(1-q)$ the gate fails and a completely depolarized state is produced. 
The average gate fidelity ${\bar F}$ for this imperfect operation is given by ${\bar F}=(3q+1)/4$, which implies an error rate $p=3/4(1-q)$. Such an error model may be used to reflect our restricted knowledge on the kind of error. Imperfect projective measurements are described by the positive operators
${\cal P}_{A_2}^{(0)}= \eta |0\r_{A_2}\l 0|+ (1-\eta) |1\r_{A_2}\l 1|,{\cal P}_{A_2}^{(1)}= \eta |1\r_{A_2}\l 1|+ (1-\eta) |0\r_{A_2}\l 0|,$ 
where e.g. in the case of $\rho=|0\r\l 0|$ the correct measurement outcome, $'0'$, is obtained with probability $\eta$.
For simplicity, we consider $\eta=q$. Our analysis considers both (nested) entanglement purification with imperfect means as well as realization of the logical two--particle gate using noisy entangled states and imperfect single--particle operations. Fig. (\ref{Fig1}) shows the achievable gate error rate for logical two--qubit operations as a function of single--particle gate error rate. The different curves correspond to different numbers of nesting levels for entanglement pumping and different gate fidelities of physical two--particle interaction. Single--particle operations with a low error rate allow us to decrease error rates of two--particle operations by several orders of magnitude.

In order to realize the scheme, it is necessary that the operations we perform respect the virtual tensor-product structure we imposed. In particular, our scheme requires the realizability of the following operations: 
(i) Entangling two--particle gate ${\cal E}$ acting only on specific virtual subsystems of each particle, e.g. $A_2$, $B_2$, without affecting other virtual subsystems.
(ii) Single--particle measurement on one virtual subsystem without affecting other virtual subsystems.
(iii) Arbitrary unitary operations on one virtual subsystem.
(iv) CNOT and SWAP gates \cite{noteCNOT} between arbitrary virtual subsystems, i.e. CNOT$_{A_jA_k}$, SWAP$_{A_jA_k}$.

On the one hand, (i) imposes conditions on the (non--local) two--particle interactions, namely that the completely positive map ${\cal E}$ representing the two--particle operation between virtual subsystems $A_2$ and $B_2$ should 
act as $\eins$ on the remaining virtual subsystems. Note that the division of $d$--dimensional Hilbert space into virtual subsystems is arbitrary and may we choosen in such a way that this condition can be fulfilled. In addition, ${\cal E}$ needs to be able to create entanglement, which can e.g. be checked using the results of Ref. \cite{Ci00}. For imperfect two--particle operations which can be described by a map ${\cal E}'_{U_{A_1B_1}}$ similarly to Eq. (\ref{noisyU}) with $U_{A_1B_1}=$CNOT, the gate is entangling iff $q'>1/9$. For high fidelity single-- particle operations, this also determines the highest tolerable error rate $p\approx 2/3$ of physical two--particle gates for which our scheme is applicable. 
Conditions (ii)-(iv) concern (local) single particle operations and may be replaced by the ability to perform arbitrary single--particle operations, however this may be more difficult to achieve. While (iii) ensures the ability to manipulate the logical qubit, (ii) and (iv) are required to realize nested entanglement pumping and for the realization of the logical two--qubit gate by consuming entanglement. For example, CNOT and SWAP operations between virtual subsystems $X_2$ and $X_3$, together with ${\cal E}$, are required to create and purify two noisy pairs. The purification step also involves measurements on virtual subsystems $X_2$, which clearly should not affect other virtual subsystems that are used to store logical qubits or other entangled pairs. 

We would also like to emphasize that metastable states (i.e. long decoherence times) are solely required for the virtual subsystems $X_1$, as the additional virtual subsystems are used only when implementing a two--qubit interaction. To be specific, coherence of the additional virtual subsystems is required only on timescales needed to implement a logical two--qubit gate using the scheme described above, while coherence of logical qubits ($X_1$) has to be maintained, as usual, over the whole time required for the quantum computation. This allows one to use e.g. motional states of ions or neutral atoms as virtual subsystems which decoherence time is much shorter than the one of the electronic states. 


We will now briefly discuss possible implementations of our scheme, taking conditions (i)-(iv) into account.
As and example consider an array of ions trapped in microtraps, following the proposal of Ref. \cite{Ci02}. The electronic and motional states of trapped ions provide the additional levels required to implement our scheme. The motional states (in $x$ and $y$ directions) are used to temporally store logical qubits and previously generated entangled states, while electronic states of ions are used to generate entanglement between neighboring ions by applying the two--particle gate proposed in Ref. \cite{Ci02} which is based on Coloumb interaction. 
Requirements (i),(iii) and (iv) can be met in such a set--up using present--day technology \cite{Du02}. In fact, several ingredients, e.g. local CNOT gates between electronic and motional states have already been experimentally demonstrated \cite{Le02}. However, (ii), the measurement of electronic states using spectroscopic methods seems to require either tighter traps (smaller Lamb-Dicke parameter) or more efficient detectors. Alternatively, one may embed each ion into a cavity, thereby directing the emission of photons into $z$--direction to avoid recoil kicks in $x$ and $y$ direction which would otherwise destroy the coherence of the motional states. Similarly, neutral atoms trapped in dipole traps or optical lattices may be used, with e.g. the two--particle gate proposed in Ref. \cite{Ja00} is applied. 

Our scheme is also applicable in a {\it concatenated} scenario in situations where operations at different levels are not equally difficult to realize. In this case, operations at the lowest concatenation level completely determine the achievable quality of operations at highest level. Such a situation may e.g. occur in distributed quantum computation \cite{Ci97}. Consider several ions stored in a Paul trap, with at least one ion embedded into a cavity. Several such systems may be placed in the same lab and connected by optical fibers, while several such labs form the set--up for quantum computation. High--quality single--ion manipulations (electronic and motional state) allow us to increase the quality of interaction between ions in the same trap. This high--quality multi--ion interactions can then be used to increase the quality of interactions between ions in different traps (which typically involves the exchange of photons and is thus of very low quality and even not deterministic), and finally to increase quality of interactions between distant labs and to realize such operations deterministically. Note that in such set--up, the problem of measurements (ii) at level 2 can be avoided by using an auxiliary ion solely for measurement purposes in such a way that the electronic and motional states of other ions are not affected.

We have shown that entanglement purification is a useful tool also for quantum computation and allows one to reduce error thresholds for two--particle operations by several orders of magnitude. This in turn implies that small ($d \leq 32$), well controllable physical systems which interact (in a possible very noisy or even probabilistic way) are sufficient to build a fault--tolerant quantum computer. 
We have illustrated our proposal with trapped neutral atoms and ions. We however believe that they are applicable to other existing proposals for quantum computation (e.g. based on linear optics \cite{Kn01,Pa01}), or may even trigger the design of fault tolerant proposals especially suitable to meet the requirements of our scheme.


We would like to thank J. I. Cirac for many useful comments and valuable discussions. This work was supported by European Union through grant No. HPMF-CT-2001-01209 (W.D.), IST-1999-13021 and the Deutsche Forschungsgemeinschaft.






\begin{thebibliography}{99}
\bibitem{noteCNOT}
The CNOT [SWAP] operation is defined by the following mapping of states, written in the
standard basis: $|i\rangle_A|j\rangle_B \rightarrow |i\rangle_A|i\oplus
j\rangle_B$ [$|i\rangle_A|j\rangle_B \rightarrow |j\rangle_A|i\rangle_B$], where $\oplus$ denotes addition modulo 2.
\end{thebibliography}
\end{document}